\documentclass{aastex701}
\usepackage{amsmath}
\usepackage{natbib}
\usepackage{upgreek}

\accepted{Nov 6, 2025}
\submitjournal{AJ}

\begin{document}

\title{Radial Velocity confirmation of two binary systems detected with the pulsation timing method
}

\author[orcid=0000-0002-1691-8479,sname='Otani']{Tomomi Otani} 
\affiliation{Department of Physical Sciences, Embry-Riddle Aeronautical University, 1 Aerospace Blvd, Daytona Beach, FL 32114, USA}
\email{otanit@erau.edu}  

\author[orcid=0000-0001-6446-6617]{A.S.\,Baran}
\affiliation{ARDASTELLA Research Collaboration, Physics, Astronomy, and Material Science Department, Missouri State University, 901 S National Ave, Springfield, MO 65897, USA}
\email{AndrzejBaran@MissouriState.edu}  

\author{Kaydence Boyle} 
\affiliation{Department of Physical Sciences, Embry-Riddle Aeronautical University, 1 Aerospace Blvd, Daytona Beach, FL 32114, USA}
\email{boylek2@my.erau.edu} 

\author{Julia M. Clark} 
\affiliation{Department of Physical Sciences, Embry-Riddle Aeronautical University, 1 Aerospace Blvd, Daytona Beach, FL 32114, USA}
\email{CLARJ117@my.erau.edu}

\begin{abstract}
We present follow-up radial velocity (RV) results for the subdwarf B\,+\,white dwarf binaries TIC\,202354658 and TIC\,69298924, which were originally identified using the pulsation-timing and Fourier-modulation methods applied to ultra-short-cadence photometric data from the Transiting Exoplanet Survey Satellite (\textit{TESS}). Using spectra obtained with the Gemini-North telescope and the Nordic Optical Telescopes, we derived the radial velocity curves for both targets. The orbital solutions we obtained are consistent with the predictions based on pulsation timing, and they provide more precise orbital information compared with the previously obtained results. We therefore confirm that both TIC\,202354658 and TIC\,69298924 are binary systems with white dwarf companions. The RV solution of TIC\,202354658 indicates a relatively large eccentricity (e = 0.313 ± 0.013), which would be uncommon for a short orbital period sdB binary.
\end{abstract}



\section{Introduction} \label{sec:introduction}

Subdwarf B (sdB) stars are extreme horizontal branch stars known to be helium-burning and to retain a residual hydrogen envelope on top of a helium shell. Binary population synthesis (BPS) models, constructed by \citet{han02,han03}, are among the most extensively studied models of subdwarf B star evolution. These models indicate that hot subdwarfs must have undergone binary evolution, including the common envelope (CE) phase, Roche lobe overflow (RLOF), or white dwarf mergers.

Approximately 30\% of sdB stars show detectable pulsations. Although pulsation amplitudes have been found to vary \citep[e.g.][]{kilkenny10}, the pulsation periods of subdwarf B variable (sdBV) stars are usually stable \citep[e.g.][]{ostensen01}, and they can therefore be considered good chronometers. This allows us to use the pulsation timing method to detect binary motion and to obtain orbital solutions of the systems.


The pulsation timing method is suitable for detecting the orbital motion of longer-period binary systems. Changes in a star's radial distance due to binary motion induce periodic shifts in light travel time. These changes can be measured via pulsation timing \citep{otani22}. Detecting these changes is particularly efficient when continuous photometric observations, such as those from {\it Kepler} and {\it TESS}, are available. 
The periodic change in pulse arrival times can also be measured using the frequency modulation (FM) method \citep{shibahashi12}. This change in pulse arrival times can be readily obtained from a Fourier analysis of the light curves, making this method particularly useful when continuous data are available.

\citet{otani25} used the pulsation timing and FM methods to investigate the \textit{TESS} light curves of 10 sdB stars and found that two of them (TIC\,202354658 and TIC\,69298924) are likely to have white dwarf companions with orbital periods of 5.5 d and 12.8 d, respectively.

Confirming binary systems using other techniques like the radial velocity (RV) method is essential for both the pulsation timing and the FM methods, as it provides an independent confirmation of binary motion. So far, only two sdB systems, CS\,1246 \citep{barlow11b} and BPM\,36430 \citep{smith22}, have been detected by the pulsation timing method and confirmed by the RV method. Here, we present our RV analysis for two sdB binary candidates (TIC\,202354658 and TIC\,69298924) previously reported by \citet{otani25}.

Section~\ref{sec:observation_and_method} describes the facilities and instruments used to obtain spectroscopic data and their reduction procedure. Section~\ref{sec:results} presents our analysis of the RV data. We discuss the results and present our conclusion in Section~\ref{sec:discussion}, while Section~\ref{sec:summary} provides a brief summary of our work. 

\section{Observations and method} \label{sec:observation_and_method}

We obtained spectroscopic data of TIC\,69298924 in queue mode, with the Gemini Multi-Object Spectrograph (GMOS) attached to the Gemini North telescope \citep{hook04}. A total of nine usable spectra were collected between September 22nd and October 5th, 2025 (Program ID: GN-2025B-Q-306, PI: T. Otani). We used the B480 grating centered at 4500\,\AA, a 1.0-arcsec slit, allowing for a spectral resolution of 6.5\,\AA, and 1\,$\times$\,4 binning. The spectral resolution Each spectrum had an exposure time of 289\,s. A flat-field frame and a CuAr lamp spectrum were taken after each exposure. To process the collected data, we used the \texttt{GMOS} package within the Image Reduction and Analysis Facility (\texttt{IRAF}) \citep{tody86, tody93}. The signal to noise ratio (S/N) of individual spectra is about 110.

Spectroscopic data of TIC\,202354658 were collected with the Alhambra Faint Object Spectrograph and Camera (ALFOSC) mounted on the 2.56\,--\,m Nordic Optical Telescope (NOT) at the Roque de los Muchachos Observatory on La Palma. A total of 11 usable spectra were collected between July 13th and August 13th, 2025 (Program ID: 71-408, PI: T.Otani, co-I: A.S.Baran). We used ALFOSC with grism \#18 and a 0.5" slit allowing for a spectral resolution of 2.2\,\AA, and a 900\,s exposure time. A flat-field frame and a He lamp spectrum were also acquired. To process the collected data, we used standard tasks within \texttt{IRAF}. The S/N of individual spectra is about 80.

In both cases, the star’s spectral lines were cross-correlated against a template spectrum using the \texttt{IRAF} \texttt{fxcor} task, which gives each radial velocity along with a formal error from the height and width of the correlation peak.
 
\section{Results} \label{sec:results}

The light travel time delay, $\uptau$ can be described as $\uptau$\,=\,z/c, where {\it z} is the radial displacement of a star relative to the center of mass of the binary system, measured along the line of sight to an observer. We assume a direction away from the observer to be positive. The RV is the time derivative of the radial displacement, {\it z}, therefore, if the orbit were circular,  the RV curve can be expressed as a sine curve with the same period but a phase difference of $\uppi$/2 in phase angle compared to the pulsation timing method. If the orbit were eccentric, the phase difference would not be exactly $\uppi$/2, but their phases would still be different.

The RV amplitude can also be determined from the pulsation timing method, 

\begin{equation}
{\rm K_{sdB} = \frac{2\,\pi\,c\,A_{\uptau}}{P_{orb}\sqrt{1-e^2}}}
\label{eq:K1}
\end{equation}

where A$_{\uptau}$ is the amplitude of the pulsation timing variation, P$_{\rm orb}$ is the orbital period of the binary system,  {\it c} is the speed of light, and {\it e} is the eccentricity. 

In the following subsections, we present the results of our RV analysis. We used a binary model \citep[Equation 65 of][]{MurrayCorreia2010} to obtain an orbital solution. The number of RV data points is smaller than the number of the pulsation timing data points in \citet{otani25}, therefore we fit the former with the fixed orbital period, which was reported by \citet{otani25}. Since the RV curve is more sensitive to eccentricity, we allowed the eccentricity ($e$) and the argument of periastron ($\upomega$) to vary in the RV fits. Next, we refit the pulsation timing variation data in \citet{otani25} to obtain the updated orbital parameters. Here, we used fixed values for $e$ and $\upomega$ which were derived from the previous steps. For the pulsation timing variation fitting, we used Equation (12) of \citet{otani22} to obtain orbital parameters.

\subsection{TIC\,69298924} \label{sec:rv_results_TIC6929}
\begin{figure} 
\plotone{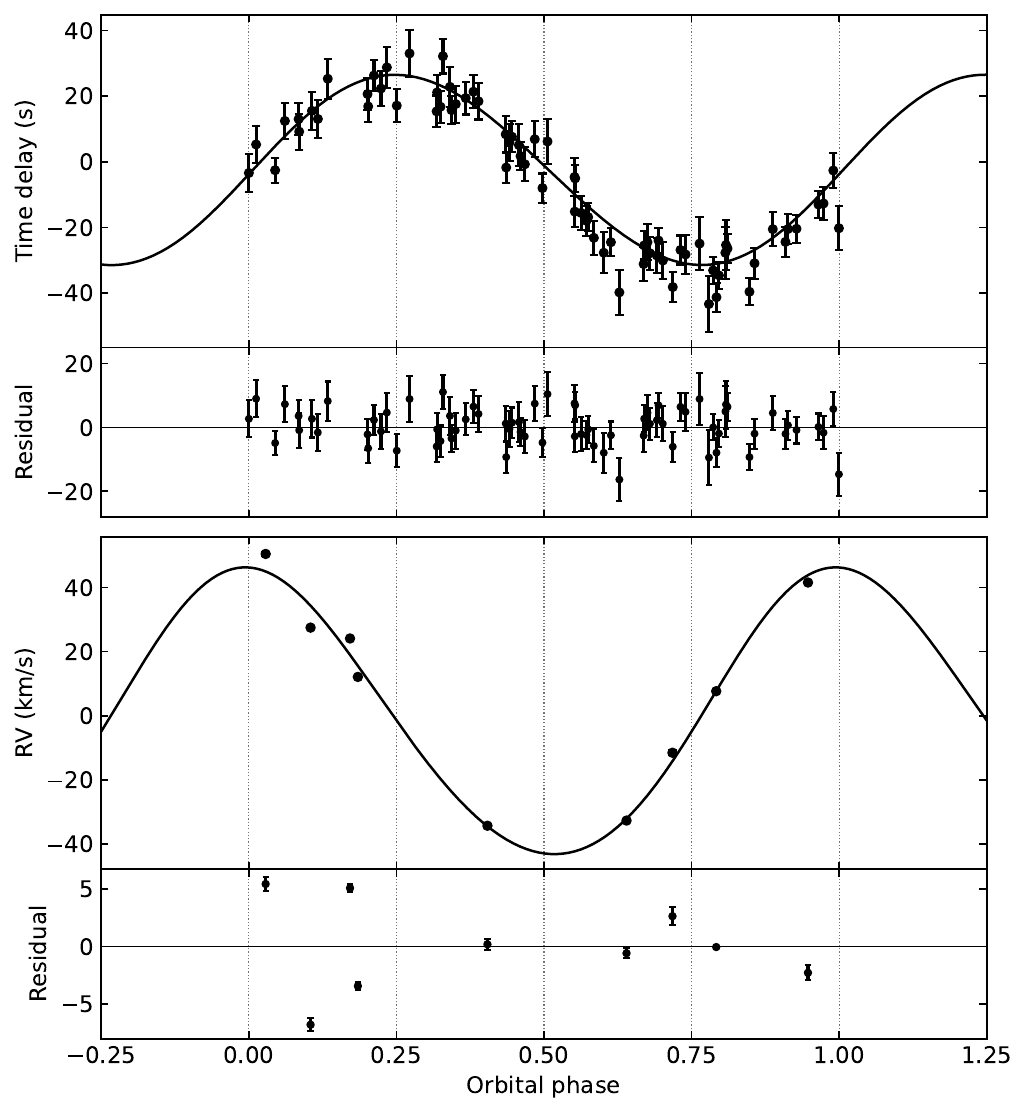} 
\caption{Comparison of the pulsation timing and the RV variations for TIC\,69298924. The pulsation timing variation is shown in the top panel, while the RV variation is shown in the bottom panel. Solid curves represent the eccentric-orbit fits. The bottom part of each panel shows the residuals. Uncertainties of the RV data points are smaller than the size of the symbol.}
\label{fig:rv_TIC6929} 
\end{figure}

The target information is listed in Table\,\ref{tab:target}.  The RV data points and their uncertainties used in our analysis are listed in Table\,\ref{tab:RV_TIC_6929}. In Figure\,\ref{fig:rv_TIC6929}, we show both the pulsation timing data (top panel) and the RV data (bottom panel). The same reference time, BJD$_{\rm TDB}$\,=\,2459809.396615, was applied to both data sets. The pulsation timing data were taken from \citet{otani25} and phased using their orbital period of 12.83569\,days. The solid curve in the upper part of the top panel represents the eccentric\,--\,orbit fit. The lower part of this panel shows residuals. The upper part of the lower panel shows phased RV data points. These data points were phased with the orbital period given above. The solid curve represents the eccentric\,--\,orbit fit. The residuals are shown in the lower part of this panel. Since the eccentricity of the orbit, which was obtained from the RV method, is close to 0, the phase shift between the maxima of the pulsation timing and RV variations is approximately 0.25 ($\pi/2$ in phase angle).

Table~\ref{tab:fm_vs_oc} lists the RMS and $\chi ^2$ of the residuals for the pulsation-timing and RV solutions. The pulsation timing solution provides a reasonable fit; however, the RV residuals are larger than their uncertainties. Since pulsation timings analysis shows a reasonable fit, the excess RV scatter may not be due to an orbital model but due to some effects not included in the uncertainty, like stellar pulsations and instrumental systematics. Stellar rotation may also contribute, but not a large portion. Because of those reasons, the uncertainties of the orbital parameters derived from the RV method may be underestimated.


Orbital solutions for both fits are given in Table \,\ref{tab:fm_vs_oc}. This table also provides a comparison of the FM, pulsation timing, and RV results. In the case of the FM method, we used the technique described in \citet{shibahashi12} to calculate the orbital period (P$_{\rm orb}$) and radial velocity amplitude (K$_{\rm sdB}$). We used frequencies and amplitudes reported by \citet{baran23}, who analyzed \textit{TESS} ultra short cadence data collected in Sectors\,44\,--\,46. The pulsation timing method results are taken from \citet{otani25} using the light curves of Sectors\,44\,--\,46 and 71\,--\,72 of the \textit{TESS} ultra\,--\,short cadence data. The expected amplitude of the RV variation, which was obtained using the amplitude of the time delay in the upper panel (A$_\uptau$\,=\,29.1\,sec) and equation\,\ref{eq:K1}, is K$_{\rm sdB}$\,=\,49.4\,km/s \citep[also reported by ][]{otani25}. The amplitude derived from the RV fit is about 10\% smaller than the results obtained from the pulsation timing method. The limited number of RV data points is likely responsible for this discrepancy. Given the results described above, we confirm the sdB star has a companion, as first reported by \citet{otani25}.

\startlongtable
\begin{deluxetable}{ccccccccc}
\tablecaption{Target information. \label{tab:target}}
\tablehead{\colhead{Name} & \colhead{RA (J2016.0)} & \colhead{DEC (J2016.0)} & \colhead{parallax} & \colhead{pmra} & \colhead{pmdec} & \colhead{RUWE} & \colhead{G} & \colhead{spectral types}\\ 
 & \colhead{(degrees)} & \colhead{(degrees)} & \colhead{(mas)} & \colhead{(mas yr$^{-1}$)} & \colhead{(mas yr$^{-1}$)} &  & \colhead{(mag)} & } 

\startdata
TIC\,69298924 & 121.7362 & 15.4555 & 0.67 $\pm$ 0.04 & 0.45 $\pm$ 0.04 & -6.14 $\pm$ 0.03 & 1.04 & 14.75  & sdB\\
TIC\,202354658 & 236.2900 & 59.9181 & 0.81 $\pm$ 0.03 & -12.56 $\pm$ 0.04 & 8.76 $\pm$ 0.04 & 1.09 & 14.41 & sdB\\
\enddata
    \tablecomments{The RA, DEC, parallax, pmra, pmdec, RUWE, G magnitude were obtained from the Gaia archive \citep{GAIADR3_23}. RUWE was used to assess the quality of the single-source astrometric solution \citep{lindegren18}}
\end{deluxetable}

\startlongtable
\begin{deluxetable}{ccc}
\tablecaption{RV measurements for TIC 69298924. \label{tab:RV_TIC_6929}}
\tablehead{\colhead{Time} & \colhead{RV} & \colhead{$\sigma_{\rm RV}$}\\ 
\colhead{(BJD-2457000)} & \colhead{(km/s)} & \colhead{(km/s)} } 

\startdata
3941.1388 & 24.08 & 0.34\\
3944.1258 & -34.25 & 0.48\\
3947.1503 & -32.65 & 0.44\\
3948.1490 & -11.52 & 0.78\\
3949.1003 & 7.66 & 0.20\\
3951.0948 & 41.50 & 0.67\\
3952.1401 & 50.41 & 0.62\\
3953.1169 & 27.45 & 0.55\\
3954.1452 & 12.11 & 0.32\\
\enddata
    \tablecomments{Timestamps are given for mid-exposures.}
    \tablecomments{Table\,\ref{tab:RV_TIC_6929} is published in its entirety in the machine-readable format.}
\end{deluxetable}



\subsection{TIC 202354658} \label{sec:rv_results_TIC2023}
\begin{figure}[] 
\plotone{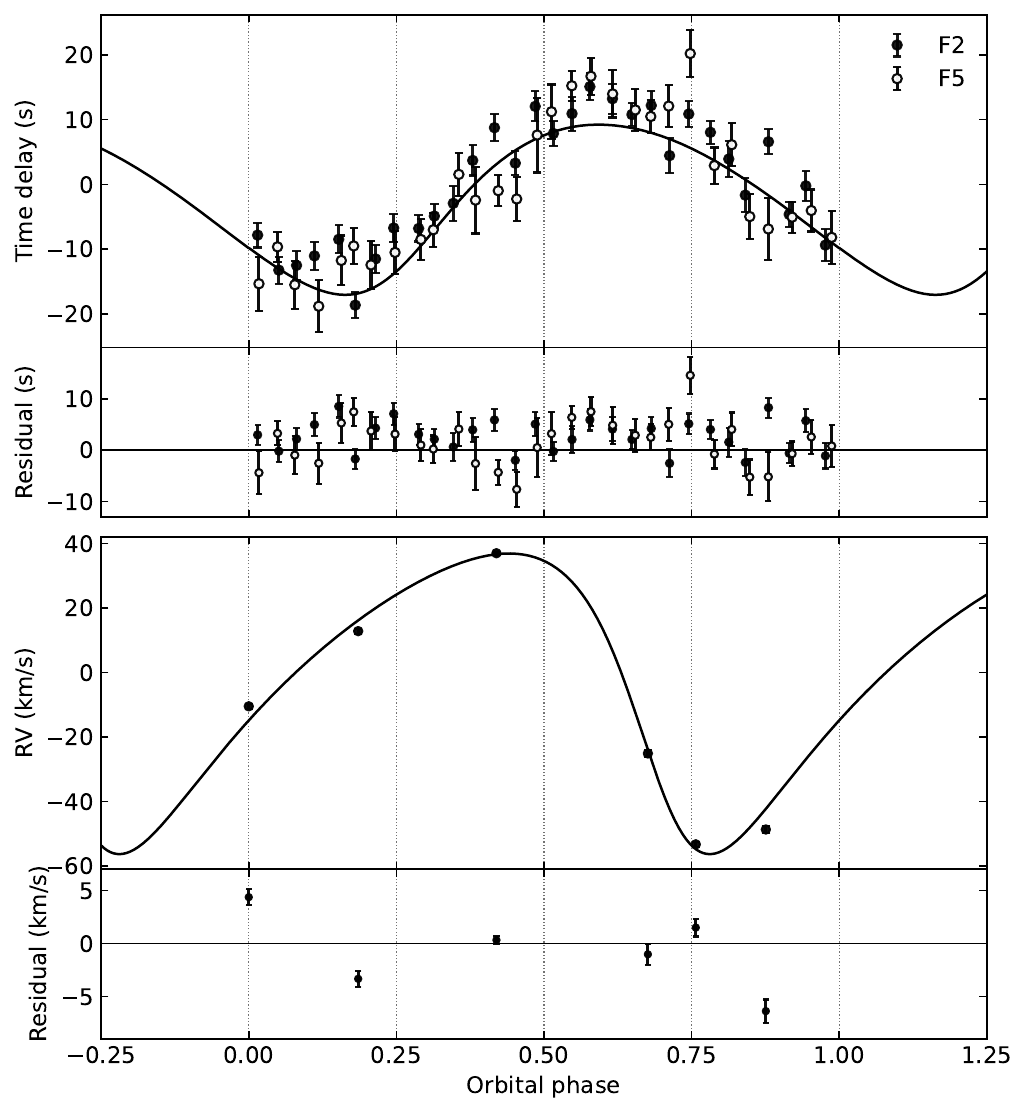} 
\caption{Comparison of the pulsation timing and the RV variations for TIC\,202354658. The pulsation timing variation is shown in the top panel, while the RV variation is shown in the bottom panel. Solid curves represent the eccentric orbit fits. Bottom parts of each panel show residuals of the fit. Uncertainties of the RV data points are smaller than the size of the symbol.}
\label{fig:rv_TIC2023} 
\end{figure}

The target information is listed in Table\,\ref{tab:target}.  The RV data points and their uncertainties are listed in Table\,\ref{tab:RV_TIC_2023}. In Figure\,\ref{fig:rv_TIC2023}, we show a comparison of the pulsation timing (top panel) and the RV (lower panel) methods. The same reference time, BJD$_{\rm TDB}$\,=\,2460128.474813, was applied to both data sets. The pulsation timing data were taken from \citet{otani25} and phased using their orbital period of 5.54078\,d. The data points for f2 and f5 from \citet{otani25} were binned into 30 phase bins for each frequency. All binned data points and uncertainties were calculated using weighted means. The solid curve in the upper part of the top panel represents the eccentric\,--\,orbit fit. The lower part of the same panel shows residuals. The upper part of the bottom panel shows the phased RV data points. These data points were phased using the orbital period given above. The solid curve represents the eccentric\,--\,orbit fit. The residuals are shown in the lower part of this panel. Since the eccentricity of the orbit, which was obtained from the RV method, is far from 0, the phase shift between the maxima of pulsation timing and the RV variations is different from 0.25 ($\pi/2$ in phase angle).

Table~\ref{tab:fm_vs_oc} lists the RMS and $\chi ^2$ of the residuals for the pulsation-timing and RV solutions. The pulsation timing solution provides a reasonable fit; however, the RV residuals are larger than their uncertainties. Since pulsation timings analysis shows a reasonable fit, the excess RV scatter may not be due to an orbital model but due to some effects not included in the uncertainty, like stellar pulsations and instrumental systematics. Stellar rotation may also contribute, but not a large portion. \citet{Baran24} and \citet{otani25} indicate TIC\,202354658 may be rotating with $P_{rot}$ = 17.3 d. If we assume the radius of the star is a typical sdB star radius, R = 0.20 $R_\odot$, $v \sin{i}$ $\leq$  0.6 km/s. Because of those reasons, the uncertainties of the orbital parameters derived from the RV method may be underestimated.



Orbital solutions for both fits are also given in Table~\ref{tab:fm_vs_oc}. This table also shows a comparison of the FM, pulsation timing, and RV results. For the FM method, we used two multiplets, ($f_1$, $f_2$, $f_3$) and ($f_4$, $f_5$, $f_8$), reported by
\citet{Baran24}, who analyzed in \textit{TESS} ultra-short-cadence data from Sectors~48--51. The values in Table~\ref{tab:fm_vs_oc} are weighted means of the results obtained from both multiplets. The expected RV amplitude, which was obtained from the amplitude of the time delay in the upper panel (A$_\uptau$\,=\,13.48\,s) and equation~\ref{eq:K1}, is K$_{\rm sdB}$\,=\,55.8\,km\,s$^{-1}$ \citep[also reported by ][]{otani25}. The RV amplitude of the fit for this target is about 15\% smaller than the result obtained from the pulsation timing method. Again, we suspect that the limited number of RV data points is the cause. Given the results described above, we confirm that the sdB star has a companion, as first reported by \citet{otani25}.

\startlongtable
\begin{deluxetable}{cccc}
\tablecaption{Radial Velocity data of TIC 202354658. \label{tab:RV_TIC_2023}}
\tablehead{\colhead{Time} & \colhead{RV} & \colhead{$\sigma$ RV} & \colhead{comment}\\ 
\colhead{(BJD-2457000)} & \colhead{(km/s)} & \colhead{(km/s)} & } 

\startdata
3869.5268 & -10.47 & 0.74 & \\
3870.5535 & 12.84 & 0.76& \\
3877.4615 & 47.5 & 1.2& *1\\
3883.4017 & 35.3 & 1.2& \\
3885.4610 & -48.7 & 1.1& \\
3888.4342 & 39.11 & 0.55& *1\\
3888.4696 & 26.3 & 1.0& *1\\
3894.3993 & 44.4 & 1.3& *1\\
3895.3928 & -29.4 & 1.6& *2\\
3895.4636 & -22.2 & 1.3& *2\\
3899.4060 & 27.4 & 1.1& *1\\
3901.4259 & -53.27 & 0.84& \\
\enddata
    \tablecomments{Timestamps are given for mid-exposures.}
    \tablecomments{Table\,\ref{tab:RV_TIC_2023} is published in its entirety in the machine-readable format.}

    \tablecomments{For *1, we used the weighted mean of those five points taken in a similar phase. }

    \tablecomments{For *2, we used the weighted mean of those two points taken in a similar phase. }

\end{deluxetable}

\begin{deluxetable}{lclr}[ht!]
\tablecaption{Orbital solutions derived from three methods, with goodness-of-fit statistics for the pulsation-timing and RV solutions \label{tab:fm_vs_oc}}
\tablewidth{0pt}
\tablehead{
\colhead{\textbf{Parameter}} & \colhead{\textbf{FM Method}} & \colhead{\textbf{Pulsation Timing Method}} & \colhead{\textbf{RV Method}}
}
\startdata
\multicolumn{4}{c}{\textbf{TIC 69298924}} \\ [1ex]
\tableline
$\upgamma$ (km/s)     &  &  & -0.78 $\pm$ 0.18\\
$\tau$ (s)     &  & 29.04 $\pm$ 0.88 & \\
K$_{\rm sdB}$ (km/s)  & 50.9    $\pm$ 2.3    & 49.4     $\pm$ 1.5     &  44.66 $\pm$ 0.28 \\
P$_{\rm orb}$ (d)     & 12.928 $\pm$ 0.046 & 12.83569 (fixed) & 12.83569 (fixed)\\
$e$ & & 0.0643 (fixed)      &  0.0643 $\pm$ 0.0050\\
$\upomega$ (deg) & & 357     $\pm$ 52     &  324.7 $\pm$ 4.9\\
T$_{\rm peri}$ (BJD) & &   2459809.4 $\pm$ 1.9    &  2459808.21 $\pm$ 0.17\\
RMS$_{\rm RV}$ (km/s) & & & 3.76\\
RMS$_{\rm timing}$ (s) & & 5.72 & \\
$\chi^2$/dof                & & 79.9/68 & 601.3/4\\
\hline
\multicolumn{4}{c}{\textbf{TIC 202354658}} \\ [1ex]
\tableline
$\upgamma$ (km/s)     &  &  & 0.83 $\pm$ 0.34\\
$\tau$ (s)     &  & 13.48 $\pm$ 0.61 & \\
K$_{\rm sdB}$ (km/s)  & 59.1    $\pm$ 3.4    & 55.9     $\pm$ 2.5     & 46.61 $\pm$ 0.46\\
P$_{\rm orb}$ (d)         &   5.5421 $\pm$ 0.0090 &  5.54078 (fixed) & 5.54078 (fixed)\\
$e$ & & 0.313 (fixed)     &  0.313 $\pm$ 0.013\\
$\upomega$ (deg) & & 136     $\pm$ 16     &  127.20 $\pm$ 2.2\\
T$_{\rm peri}$ (BJD) & &   2459614.81 $\pm$ 0.58   &  2459615.664 $\pm$ 0.026\\
RMS$_{\rm RV}$ (km/s) & & & 3.52\\
RMS$_{\rm timing}$ (s) & & 4.51 & \\
$\chi^2$/dof                & & 182.57/54 & 92.62/1\\
\enddata
\tablecomments{To obtain $e$ in the FM method, we need to detect the second sidelobe \citep[see Equation (76) of ][]{shibahashi12}; however, their amplitudes were below the noise levels for both targets.}
\tablecomments{$\tau$ is the amplitude of the time delay ($\tau = ac\sin{i}$, where $a$ is the semimajor axis, $c$ is the light speed, and $i$ is the inclination of the system.)}
\label{result_comp}
\end{deluxetable}

\section{Discussion and Conclusion} \label{sec:discussion}
The two stars analyzed in this work were found to be in binary systems by \citet{otani25}. They used the pulsation timing method to demonstrate that the pulsations are modulated by orbital motion. The triplets detected in the amplitude spectra itself could not distinguish orbital modulation from rotational splitting without additional analysis. However, as pointed out in \citet{otani25}, rotational multiplets can be distinguished from orbital modulation by analyzing the amplitude behavior. The authors derived a significant variation in the time delays and no amplitude modulation of the multiplet components. These findings led \citet{otani25} to conclude that these two pulsating stars analyzed in their work reside in binary systems.

The intention of this work was to verify their discovery using an independent method, {\it i.e.} to measure the RV of the pulsating components in those binary systems. Our analysis supported the predictions based on the pulsation timing. The RV data points of both stars show periodic RV variations. These results confirmed the interpretation of TIC\,69298924 and TIC\,202354658 as binary systems. The conclusions in \citet{otani25} are further strengthened.

The TESS light curve of TIC\,202354658 was analyzed by \citet{Baran24}, who identified three multiplets of two types found in the amplitude spectrum. Two multiplets were interpreted as signatures of binary motion, while the third one was interpreted as rotationally induced. Our work confirmed that preliminary interpretation by \citet{Baran24}, specifically (f$_1$ f$_2$, f$_3$) and (f$_4$, f$_5$, f$_8$) are binary\,--\,related, which makes (f$_5$, f$_6$, and f$_7$) is likely a rotationally\,--\,induced triplet. Therefore, frequency f$_5$ seems to be a part of both binary and rotation induced multiplets.

Eccentric orbits are not commonly reported among short-period sdB binaries. \citet{edelmann05} could detect eccentricities ($e \approx$ 0.02\text{--}0.06) in five systems out of 15 systems, and the other 10 stars had eccentricity below their detection limit ($e$ = 0.02). \citet{copperwheat11} obtained the orbital eccentricities of 28 sdB binary systems and all are in the range of $e \approx$ 0.01\text{--}0.15. \citet{telting12, telting14} found systems with eccentricities 0.063 and 0.056. The eccentricity of TIC\,202354658 derived here ($e=0.313 \pm 0.013$) would be much larger than any of those short-period sdB binaries, and that makes this system interesting. However, as discussed in Sec~\ref{sec:rv_results_TIC2023}, its uncertainty may be underestimated. Therefore, the RV observations covering the full orbital phase are needed to confirm it.

As indicated in \citet{otani25}, the TIC\,69298924 and TIC\,202354658 systems are likely to have WD companions. According to \citet{Schaffenroth23}, the hot subdwarf systems with WD companions show a broad orbital period distribution ranging from about one hour to 27\,days, with two distinct peaks at around one day and around 5\,--\,10\,days. TIC\,202354658 falls within the latter peak, while the orbital period of TIC\,69298924 is slightly longer than the location of this peak. Only a handful of sdB+WD systems with orbital periods longer than 10\,days have been detected \citep[see Figure 3 of][]{ge22}, probably owing to the limited sensitivity of other detection methods. The pulsation timing method using {\it TESS}, {\it Vera C. Rubin} Observatory, {\it PLATO} and astrometry using {\it Gaia} should allow us to investigate this and longer orbital period range.


\section{Summary} \label{sec:summary}
We used spectroscopic data from the 8.1-m Gemini\,--\,North and the  2.56-m NOT telescopes, and confirmed two sdB binary systems (TIC\,69298924 and TIC\,202354658) using the RV method. The amplitudes of the RV variations are close to the expected values derived from the pulsation timing method reported by \citet{otani25}. According to our analysis, TIC\,69298924 has a nearly circular orbit, while TIC\,202354658 is in an eccentric orbit. Our results support the accuracy of pulsation timing using high-frequency pulsations in subdwarf B stars. Furthermore, the results demonstrate the importance for large sdB binarity surveys using the photometric data, particularly for systems with orbital periods longer than a few days, where other binary\,--\,detection methods, {\it e.g.} transits and reflection effects, are less efficient.

\begin{acknowledgments}
TO acknowledges research support from the National Aeronautics and Space Administration (NASA) under Grant No. 80NSSC24K0494. TO and KB acknowledge the support from Embry-Riddle Aeronautical University startup funding and the Faculty Innovation Research in Science and Technology (FIRST) program.

Based on observations obtained at the international Gemini Observatory, a program of NSF NOIRLab, which is managed by the Association of Universities for Research in Astronomy (AURA) under a cooperative agreement with the U.S. National Science Foundation on behalf of the Gemini Observatory partnership: the U.S. National Science Foundation (United States), National Research Council (Canada), Agencia Nacional de Investigación y Desarrollo (Chile), Ministerio de Ciencia, Tecnología e Innovación (Argentina), Ministério da Ciência, Tecnologia, Inovações e Comunicações (Brazil), and Korea Astronomy and Space Science Institute (Republic of Korea). 

Based on observations made with the Nordic Optical Telescope, owned in collaboration by the University of Turku and Aarhus University, and operated jointly by Aarhus University, the University of Turku and the University of Oslo, representing Denmark, Finland and Norway, the University of Iceland and Stockholm University at the Observatorio del Roque de los Muchachos, La Palma, Spain, of the Instituto de Astrofisica de Canarias. 

This work has made use of data from the European Space Agency (ESA) mission Gaia (https://www.cosmos.esa.int/gaia), processed by the Gaia Data Processing and Analysis Consortium (DPAC, https://www.cosmos.esa.int/web/gaia/dpac/consortium). Funding for the DPAC has been provided by national institutions, in particular the institutions participating in the Gaia Multilateral Agreement.

\end{acknowledgments}

\begin{contribution}

TO was responsible for obtaining data, performed the pulsation timing analysis and the comparison of RV methods, verified the overall analysis, produced the phase diagrams, an wrote and submitted the manuscript. ASB was responsible for the spectral analysis, including RV derivation, provided suggestions on the overall analysis, and edited the manuscript. KB and TO performed the pulsation timing and RV analysis, and JMC and TO performed the FM analysis.



\end{contribution}

%
\facilities{Gemini-N, NOT}

\software{IRAF, Python, NumPy, SciPy, Matplotlib}


\bibliography{myrefs}{}
\bibliographystyle{aasjournalv7}



\end{document}